\documentclass[runningheads]{svmult}

\usepackage{makeidx}   % allows index generation
\usepackage{graphicx}  % standard LaTeX graphics tool
\usepackage{subeqnar}  % subnumbers individual equations
\usepackage{multicol}  % used for the two-column index
\usepackage{physprbb}  % modified textarea for proceedings,
\makeindex             % used for the subject index
\newcommand{\greeksym}[1]{{\usefont{U}{psy}{m}{n}#1}}
\newcommand{\umu}{\mbox{\greeksym{m}}}

\begin{document}
\title*{The History of Starburst Galaxies} 
\toctitle{The History of Starburst Galaxies}
% allows explicit linebreak for the table of content
%
%
\titlerunning{Starburst History}
% allows abbreviation of title, if the full title is too long
% to fit in the running head
%
\author{Andrew W. Blain}
\authorrunning{Andrew W. Blain}
% if there are more than two authors,
% please abbreviate author list for running head
%
%
\institute{Institute of Astronomy, Madingley Road, Cambridge, CB3 0HA, UK} 

\maketitle              % typesets the title of the contribution

\begin{abstract}
Dusty 
galaxies with luminosities in excess of $10^{11}$\,L$_\odot$ have been 
detected out to redshifts $z \sim 1$ by the {\it Infrared Space 
Observatory (ISO)}, and to higher redshifts using millimetre(mm)- and 
submm-wave 
cameras on ground-based telescopes. The integrated properties of these more 
distant galaxies are also constrained by measurements of the 
intensity of the submm-wave background radiation. While it is generally unclear 
whether their energy is released by gravitational accretion or by star formation, 
circumstantial evidence favours star formation. Unless these high-redshift 
galaxies are extremely massive, which is not expected from standard models of 
galaxy evolution, this luminosity cannot be sustained for more than a 
fraction of a Hubble time, and so they are undergoing some sort of `bursting' behaviour. The 
interpretation and analysis of this population is discussed, and the key observations 
for deriving a robust history of their evolution, which is likely to be the history 
of starburst activity, are highlighted. 
\end{abstract}

\section{The Evolution of Luminous Dusty Galaxies} 
From observations of low-redshift dusty galaxies using the {\it IRAS} 
satellite \cite{SandN} close to the peak of their restframe spectral 
energy distributions (SEDs), it is known that a similar amount of energy in 
the local Universe is produced by stars in dust-enshrouded and dust-free 
environments. The comoving luminosity density of dusty galaxies is also known to 
evolve strongly, from the slope of the faint counts of {\it IRAS} galaxies at 
60\,$\umu$m \cite{Bertin}, which provide information to $z \simeq 0.2$. {\it ISO} 
observations at both shorter \cite{Elbaz} and longer \cite{Juvela} wavelengths 
confirm that strong evolution continues to $z \sim 1$.  

At longer wavelengths, the redshifted emission from very luminous, high-redshift 
dusty galaxies can be detected in the mm and submm wavebands. Independent 
surveys made using the 450/850-$\umu$m SCUBA camera at the JCMT 
\cite{Smail_catalogue} have determined the counts of high-redshift dusty 
galaxies. 1.2-mm surveys using the MAMBO detector array at the IRAM 30-m 
telescope \cite{Carilli} have detected a similar population of galaxies. In three 
cases, the detection of CO emission from gas located at the position and 
redshift of a suspected optical identification (at $z=1.06$, 2.55 and 2.80)
\cite{Frayer} provides absolute confirmation of the identification. Extremely 
deep VLA radio images of the survey fields can be used to impose constraints on 
the redshifts and SEDs of the detected galaxies \cite{CandY,SIOBK}. It is likely that the 
detected galaxies are at $\bar z \simeq 2-3$, and there are very few plausible 
low-redshift ($z \le 1$) counterparts. The counts and redshift distributions of 
these distant dusty galaxies can be used to constrain models of galaxy evolution 
at high redshifts. 

In addition to the detection of individual submm-selected galaxies, 
the background radiation intensity in the mm and submm 
wavebands \cite{Fixsen} traces the integrated emission from the entire 
population of dusty high-redshift galaxies. The mm/submm-wave background 
spectrum has the form $I_\nu \propto \nu^{2.64}$, and there is no 
clear spectral break down to 2000\,$\umu$m. The lack of a break 
to a steeper slope supports the idea that high-redshift galaxies with 
redshifted SEDs peaking at about 1000\,$\umu$m are still contributing 
to the background intensity, indicating a high maximum redshift of the 
population of about 10. If the shape of the SED is assumed not to evolve 
significantly, then the luminosity density must evolve as 
$\rho_{\rm L} \propto (1+z)^{\simeq -1.1}$ \cite{BL93b,BJ} for $z \gg 1 $ 
in order to generate this background spectrum, a result which is 
independent of cosmology. Such a gently declining high-redshift luminosity 
density is naturally consistent with the incomplete redshift distributions 
of galaxies detected in mm/submm-wave surveys.

Note that any model of the evolution of dusty galaxies must 
predict the redshift distribution of submm-detected galaxies 
correctly. Submm-wave surveys are very sensitive to high-redshift 
galaxies \cite{BL93a}, and it is easy to propose models that fit both the 
observed submm-wave counts and background radiation spectrum at the 
expense of a redshift distribution that is biased far too high. The verification 
of predicted redshift distributions is therefore a crucial test of such models.

In this paper, the forms of evolution of galaxies that were previously 
derived in the context of far-infrared(IR) and submm-wave data are updated 
to take account of the much greater amount of information that has become 
available, particularly from {\it ISO} surveys. The results are similar to, but 
less uncertain than, those derived earlier \cite{BSIK,BJ}. The results have also 
been updated to include the currently favoured 
cosmological parameters $\Omega_0=0.3$, $\Omega_\Lambda=0.7$ and 
$H_0=65$\,km\,s$^{-1}$\,Mpc$^{-1}$ are assumed.

\section{Constraining the Evolution of Dusty Galaxies} 

\subsubsection{A Baseline at Low Redshifts} 

The luminosity function of {\it IRAS} galaxies is best constrained at 
60\,$\umu$m \cite{Saunders}; and information about the same population 
of galaxies is also available at 100\,$\umu$m \cite{SandN}. These wavelengths 
are close to the peak of the SED of a nearby ($z \simeq 0$) dusty galaxy 
for any reasonable dust temperature. The ratio of the bright counts at 60 and 
100\,$\umu$m imply a luminosity-averaged dust temperature 
$T \simeq 35$--45\,K. 850-$\umu$m observations of galaxies detected 
by {\it IRAS} \cite{Dunne}, with a long wavelength baseline to provide an 
excellent probe of the SED, indicate that $T = 36 \pm 5$\,K and the Rayleigh--Jeans 
spectral index is $3.3 \pm 0.2$. The population of low-redshift dusty galaxies can 
be divided into relatively short-lived warm interacting/starbursting galaxies and 
long-lived cooler quiescent galaxies \cite{BandB,BJ}; however, the details of this 
distinction are relatively unimportant for studies of high-redshift galaxy 
evolution. Any low-luminosity, low-temperature dusty galaxies missing from 
existing surveys do not contribute significantly to the luminosity 
density, even at low and moderate redshifts. 

The form of evolution of the baseline low-redshift far-IR luminosity 
function $\Phi_0(L)$ must be dominated by pure luminosity 
evolution, that is $\Phi(L,z) \simeq \Phi_0[L/g(z),0]$, to ensure that the submm 
galaxy counts and the background radiation intensity are both predicted 
correctly. Number density evolution is certainly also likely to be involved, but must be 
dominated by luminosity evolution \cite{BSIK}. The evolution function 
$g(z)$ is determined by demanding that the background radiation intensity, 
counts and redshift distributions of dusty galaxies are all in agreement 
with observations. These observations are, in order of increasing redshift, the faintest 
counts and redshift distributions of 60-$\umu$m {\it IRAS} galaxies, deep 90- 
and 170-$\umu$m counts from {\it ISO}, the spectrum of background radiation from 
{\it COBE}, and the faint counts and limited redshift information of distant 
galaxies detected using SCUBA at 450 and 850\,$\umu$m and MAMBO at 
1.25\,mm.

Several approaches can be taken to investigate the evolution. The simplest 
is to assume a form for $\Phi$, which implicitly includes details of all the physical 
processes taking place in galaxies, but is fitted to the data without 
investigating the processes in detail \cite{BSIK}. This has the advantage of requiring 
few parameters to model the galaxy SED and the form of evolution, in fact fewer 
than the number of constraining pieces of data. Well-motivated additions of 
greater complexity can thus be introduced to the models as the observations improve, 
without invoking parameters too numerous to constrain reliably and uniquely. A more 
physically motivated approach connects the evolving mass function of galaxies to 
the associated luminosity function using a prescription for both star formation and 
the fueling of active galactic nuclei (AGN) \cite{Cole,Guid,SPF}; however, care must be 
taken to avoid getting lost in the space of free parameters. Without an additional 
population of short-lived, very luminous galaxies, standard semi-analytical models,  
which include star formation in the gas that cools in galaxy disks, fail to account for 
the observed surface density of SCUBA and MAMBO sources \cite{BJ}. 

\subsubsection{Describing Evolution with a Simple Luminosity Function}

The first results derived using this approach \cite{BSIK} followed rapidly behind 
the first results of SCUBA surveys \cite{SIB}. A low-redshift 60-$\umu$m 
luminosity function was assumed \cite{Saunders}, an SED was defined by a dust 
temperature $T$, and a form of low-redshift evolution $g(z) = (1+z)^\gamma$
was included. $T$ and $\gamma$ were determined by requiring that the form of 
the 60-$\umu$m {\it IRAS} counts and the early results of deep 175-$\umu$m 
{\it ISO} surveys were reproduced; $T = 38 \pm 4$\,K and $\gamma = 3.9 \pm 0.2$ 
were required \cite{BSIK}. As discussed above, this temperature is consistent 
with subsequent SCUBA measurements of dust temperatures for {\it IRAS} 
galaxies \cite{Dunne}, while $\gamma$ matched the value inferred from optical 
surveys \cite{Lilly}, rather than the value of $\gamma = 3$ that is often assumed 
to describe the evolution of galaxies in the far-IR waveband. Note that 
$\gamma \simeq 4.5$ is derived from 15-$\umu$m {\it ISO} surveys, taking into 
account the complex restframe SED of a dusty galaxy 
between 5 and 10\,$\umu$m \cite{Xu}. 

It is now possible to use the much more extensive data from 90- and 175-$\umu$m 
{\it ISO} counts \cite{Juvela}, and a more popular non-zero-$\Lambda$ cosmology 
to revisit the results. There are no substantial changes; formally $T = 37 \pm 3$\,K 
and $\gamma = 4.05 \pm 0.15$ are the latest results, if a Rayleigh--Jeans spectral index 
of 3.5 is assumed. 

The form of evolution at higher redshifts, too distant for {\it ISO} observations, is 
constrained by the background radiation intensity \cite{Fixsen} and the 
counts of SCUBA galaxies. From 2002, mid-IR {\it SIRTF} observations should 
make a major impact in this area. Constraints from the background and SCUBA 
data are rather degenerate, although even the first SCUBA data in 1997 \cite{SIB} 
provided the tighter constraint: see \cite{BSIK}, in which various models 
of high-redshift evolution were considered. The so-called Gaussian model, in which 
$g(z)$ is represented by a Gaussian in cosmic epoch, provided the best description 
of the redshift distribution of SCUBA galaxies. The more accurate 175- and 
850-$\umu$m counts now available provide some additional information, and 
are useful for updating the results. Progress has also been made in developing a 
more appropriate form of $g(z)$, which is fully compatible with models of cosmic 
chemical evolution, and naturally includes a peak in the evolution function 
\cite{Jameson,L2000}: 
\begin{equation}
g(z) = (1+z)^{3/2} {\rm sech}^2[ b \, {\rm ln}(1+z) - c ] \, {\rm cosh}^2c. 
\end{equation} 
At low redshifts $\gamma \simeq (3/2) + 2b \sqrt{1-{\rm sech}^2 c}$. Using all 
available observational data, the results $b=2.2 \pm 0.1$ and $c=1.84 \pm 0.1$ are 
obtained; see the solid line in Fig.\,1. These results are similar to those derived earlier 
\cite{BSIK}, but take account of revised cosmological parameters and tighter error 
bars on some of the constraining data. The model described above is consistent 
with all the observed background radiation, counts and redshift distributions of 
galaxies at wavelengths longer than 60\,$\umu$m. The dominant source of energy 
in the Universe remains the restframe far-IR radiation of starlight and AGN emission 
reprocessed by dust. 

\subsubsection{Describing Evolution Using a Model of Merging Galaxies} 

In an alternative investigation, we took a simple form of the evolution of the merger 
rate of dark-matter halos \cite{BL93b,BJ}, which adequately reproduces the results 
of recent $N$-body simulations \cite{Jenkins}. We assumed that a certain 
redshift-dependent fraction $x(z)$ of the total mass of dark and baryonic matter 
involved in mergers is converted into energy by nucleosynthesis in high-mass stars 
with an efficiency 0.007$c^2$. Note that the same formalism is appropriate for 
describing the evolution of AGN fueling events at the epochs of mergers \cite{BJ}. 
The form of evolution and normalization of $x(z)$ can be determined by a joint 
comparison with the background radiation intensity and low-redshift {\it IRAS} counts. 
Using an appropriate form of $x(z) = g(z) / (1+z)^{3/2}$ \cite{Jameson,L2000}, 
$b = 1.95 \pm 0.1$, $c = 1.6 \pm 0.1$ and $x(0) = 1.35 \times 10^{-4}$ are required 
in the standard cosmology, assuming the galaxy SED discussed above. The resulting 
history of galaxy evolution is shown by the thick dashed line in Fig.\,1.

In addition, the counts of luminous galaxies associated with merger-induced bursts 
of activity can be determined if a fraction $F$ of mergers are assumed to generate 
luminous bursts of duration $\sigma$. The product $F\sigma$, which could depend on 
redshift, is the function that controls the results. Using information derived from 
the counts of both low- and high-redshift galaxies, the form of $F\sigma(z)$ required 
to account for the observations can be determined. If the form 
$F\sigma(z) = F\sigma(0) \exp(az + bz^2 + cz^3)$ is chosen, then values of 
$F\sigma(0) = 2.4$\,Gyr, $a=-4.14$, $b=-0.56$ and $c=0.46$ are required. 

\subsubsection{Faint Radio and 15-$\umu$m Mid-IR Counts} 

Both the simple and hierarchical models account for all the current observations 
in these wavebands. If the standard form of the far-IR--radio correlation is assumed, 
with a radio spectral index of $-0.65$, then the 8.4-GHz counts brighter than 
10\,$\umu$Jy predicted in the two models are 1.05 and 0.98\,arcmin$^{-2}$, with slopes 
of $-1.4$ and $-1.3$ respectively, matching the observed count 
$N(\ge S)=(1.01 \pm 0.14) (S/10\umu{\rm Jy})^{-1.25 \pm 0.2}$ \cite{Part}. The faintest 
1.4-GHz counts \cite{Carilli2} 
are also reproduced accurately. If the mid-IR 
SED is described by a power-law $f_\nu \propto \nu^\alpha$ with $\alpha = -1.95$ at 
wavelengths shorter than the peak of the SED, then the normalization and general 
features of the deep counts of galaxies determined using {\it ISO} at 
15\,$\umu$m \cite{Elbaz} are also reproduced, including the marked 
change of slope at flux densities between 0.5 and 1\,mJy. The presence or absence of 
a PAH emission feature in the SED has little effect on the results. In the hierarchical 
model, the slope of the predicted 15-$\umu$m counts at flux densities between 
1 and 10\,mJy is steeper as compared with that in the simple model, in better 
agreement with the observations.

\section{Conclusions} 

These models, which involve a minimum number of free parameters, 
provide a reasonable description of all the counts and redshift distributions of 
dusty galaxies at both high and low redshifts. As additional data, especially more 
complete redshift distributions and very deep {\it SIRTF} mid-IR counts, become 
available, more details can be incorporated into the models to reveal further
information about the properties of evolving distant dusty starbursts/AGN. 

\begin{figure}[ht]
\begin{center}
\includegraphics[width=.65\textwidth,angle=-90]{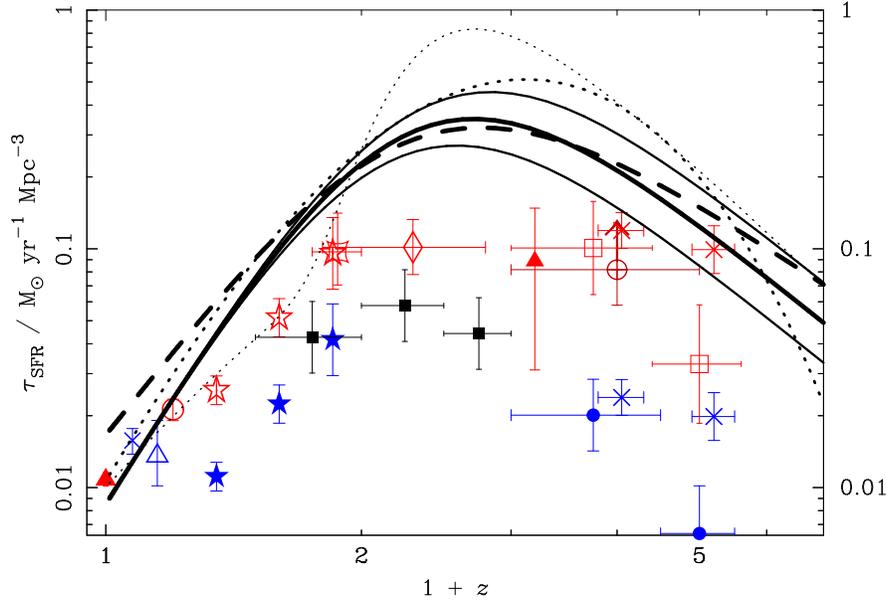}
\end{center}
\caption[]{The history of star formation inferred using both methods 
discussed in the text. The thick solid and dashed lines represent the simple 
luminosity evolution model and the hierarchical model respectively. The thinner 
solid lines show the approximate envelope of 68\% uncertainty in the results of the 
simple model. The data points are taken from a variety of sources: references can 
be found elsewhere \cite{BJ,vdW}. The thin and thick dotted lines represent the 
best-fitting results obtained in previous derivations, the modified Gaussian 
\cite{BSIK,Barger} 
and 35-K hierarchical models \cite{BJ} respectively. The absolute  
normalisation of the curves depends on the assumed stellar initial 
mass function and 
the fraction of the dust-enshrouded luminosity of galaxies that is generated by AGN. 
}
\label{eps1}
\end{figure}

\subsubsection*{Acknowledgements} 
The author, Raymond and Beverly Sackler Foundation Research Fellow at the IoA, thanks 
the Foundation for generous financial support, ESO for support at the meeting, and 
Vicki Barnard and Kate Quirk for helpful comments on the manuscript. 
These results are updated from SCUBA Lens Survey work with Ian Smail, Rob Ivison 
and Jean-Paul Kneib.

%INDEX%%%%%%%%%%%%%%%%%%%%%%%%%%%%%%%%%%%%%%%%%%%%%%%%%%%%%%%%%%%%%%%
% Please check with the editor of your book whether he plans to
% include a "mutual" subject index - if so, please code your entries
% in the standard syntax. For your own purposes you may print your
% "personal" index by using the following commands:
%
%\clearpage
%\addcontentsline{toc}{section}{Index}
%\flushbottom
%\printindex
%%%%%%%%%%%%%%%%%%%%%%%%%%%%%%%%%%%%%%%%%%%%%%%%%%%%%%%%%%%%%%%%%%%%%

\end{document}